\documentclass[aps,pre,twocolumn,superscriptaddress]{revtex4}
\usepackage{amssymb}
\usepackage{epsfig}
\usepackage{amsmath}
\usepackage{times}
\usepackage{booktabs}
\usepackage{array}
\usepackage{color}

\begin{document}
	
\title{Epidemic spreading under game-based self-quarantine behaviors: The different effects of local and global information}

\author{Zegang Huang}
\affiliation{Institute of Cyberspace Security, Zhejiang University of Technology,  Hangzhou 310023, China}
\affiliation{Binjiang Cyberspace Security Institute of ZJUT, Hangzhou 310051, China}

\author{Xincheng Shu}
\affiliation{Institute of Cyberspace Security, Zhejiang University of Technology,  Hangzhou 310023, China}
\affiliation{Binjiang Cyberspace Security Institute of ZJUT, Hangzhou 310051, China}

\author{Qi Xuan}
\affiliation{Institute of Cyberspace Security, Zhejiang University of Technology,  Hangzhou 310023, China}
\affiliation{Binjiang Cyberspace Security Institute of ZJUT, Hangzhou 310051, China}

\author{Zhongyuan Ruan}
\email{zyruan@zjut.edu.cn}
\affiliation{Institute of Cyberspace Security, Zhejiang University of Technology,  Hangzhou 310023, China}
\affiliation{Binjiang Cyberspace Security Institute of ZJUT, Hangzhou 310051, China}

\begin{abstract}
During the outbreak of an epidemic, individuals may modify their behaviors in response to external (including local and global) infection-related information. However, the difference between local and global information in influencing the spread of diseases remains inadequately explored. Here we study a simple epidemic model that incorporates the game-based self-quarantine behavior of individuals, taking into account the influence of local infection status, global disease prevalence and node heterogeneity (non-identical degree distribution). Our findings reveal that local information can effectively contain an epidemic, even with only a small proportion of individuals opting for self-quarantine. On the other hand, global information can cause infection evolution curves shaking during the declining phase of an epidemic, owing to the synchronous release of nodes with the same degree from the quarantined state. In contrast, the releasing pattern under the local information appears to be more random. This shaking phenomenon can be observed in various types of networks associated with different characteristics. Moreover, it is found that under the proposed game-epidemic framework, a disease is more difficult to spread in heterogeneous networks than in homogeneous networks, which differs from conventional epidemic models.
\end{abstract}

\maketitle\textbf{Human behaviors play a crucial role in the spread of an epidemic, and they are typically influenced by both local and global infection-related information. Nevertheless, the different effects of the two kinds of information on epidemic dynamics have not been fully explored yet. In this paper, we propose a game-based epidemic model on networks that takes into account the individual heterogeneity. We find that local information can effectively hinder the epidemic even with a small number of individuals deciding to self-quarantine. On the other hand, global information affects all individuals synchronously and can cause shaking in infection evolution curves. Significantly, our model is essentially different from conventional epidemic models in that the network heterogeneity plays a negative role in the spread of epidemics, which is contrary to the previous findings. }

\section{Introduction} 

The widespread of the COVID-19 disease in recent years has infected billions of people and led to more than $6$ million fatalities, a number that may be significantly underestimated \cite{Andy:2023}. This unprecedented global health crisis has sparked extensive research into modeling and understanding the dynamics of the epidemic \cite{Ventura:2022,Arenas:2020,Andrea:2022,Gozzi:2021,XZhang:2022,Ye:2021}. By employing traditional epidemic models, such as the susceptible-infected-removed (SIR) and susceptible-exposed-infected-removed (SEIR) models, researchers have gained valuable insights of how the coronavirus disease diffuses in the population. Nevertheless, observations from real epidemic data  demonstrate that the spread pattern of COVID-19 deviates significantly from these idealized models, which typically exhibit a single peak and a near-symmetric decline \cite{Weitz:2020,Satorras:2015}. While in reality, the epidemic curves seem more complicated, which may present oscillations and other complex features \cite{Weitz:2020,Maier:2020,Zheng:2018,Gostiaux:2023,Bestehorn:2023}.

One main reason is that the conventional models overlook the impact of human reactions during an epidemic, such as wearing face masks, reducing social interactions and taking vaccinations, which may substantially influence the epidemic dynamics \cite{Nicola:2011,Sebastian:2010,Hai-Feng:2014,Zhang1:2020,Ruan:2012,Gosak:2021}. To take this effect into account, many disease-behavior models, assuming various microscopic rules have been proposed. One preliminary approach to capturing such impact is to modify the transmission rate in classical epidemic models (called phenomenological models \cite{Wang:2016}), as the measures taken by people can generally reduce their susceptibility \cite{Sebastian:2009}. A more sophisticated approach involves considering the decision-making processes (psychological mechanisms) of individuals, which can be characterized by game theory. These disease-behavior models, according to Funk et al. \cite{Sebastian:2010}, could be broadly categorized based on the source of information,  i.e., whether the information that triggers human behaviors is globally or locally available. 

Generally speaking, global infection information (related to the prevalence of an epidemic) is disseminated through newspapers, websites, television programs, and other media channels, therefore is accessible to everyone. Several early studies have modeled this effect on epidemic spreading under the assumption of a homogeneous mixing population, where individuals interact randomly with one another \cite{Chen:2006,Onofrio:2007,Reluga:2006,Kabir:2020}. For instance, to study the interplay between vaccinating behavior and disease prevalence, Bauch proposed a game dynamic model in which the payoff of non-vaccinators is proportional to the disease prevalence \cite{Bauch:2005}. A few recent works have also considered this scenario within the multiplex network framework \cite{Xia:2019,Zhang:2022}.

Unlike global information, local infection information comes from the social or spatial neighborhood of an individual \cite{Bagnoli:2007}. This concept has been successfully applied to network models, including adaptive networks, multiplex networks and higher-order networks. For instance, in adaptive networks, it is assumed that susceptible nodes may rewire their network connections to avoid contracting the disease when they become aware of connecting with infected nodes \cite{Gross:2006,Marceau:2010,Yang:2012}. In multiplex networks, two different dynamical processes, namely information diffusion and disease spreading, take place in two distinct layers: a virtual social network and a physical contact network. These processes continue in their respective layers through the interaction between nodes and their neighborhood and can influence each other \cite{Ji:2023,Granell:2013,Wang:2014,Liu:2016,Guo:2015,zwang:2019,zwang:2021}. Furthermore, the pairwise interactions among individuals in the social layer were extended to the form of higher-order interactions, potentially giving rise to some new phenomena \cite{Chang:2023,Liu:2023,Fan:2022}.

Despite these advances, the differences between local and global information in affecting epidemic dynamics, particularly under the game-based framework (a more realistic setting), have not been fully explored yet. Although some studies have delved into this issue, they were mostly based on the phenomenological models \cite{Sebastian:2009,Wu:2012}. In this paper, we extend the classical SIR model by introducing an additional state $Q$, representing self-quarantine, and assume that the susceptible nodes would make decisions whether to self-isolate or not based on a game strategy, which takes into account the local infection status of a node (local information), the overall disease prevalence (global information), and individual heterogeneity (non-identical degree distribution). We focus on how these factors may affect the evolution of an epidemic, and aim to provide some insights into real-world epidemic dynamics. 

This paper is organized as follows. In Sec. II., we introduce our model which incorporates the game-based self-quarantine behavior into the classical SIR model. In Sec. III, we present the numerical simulations on synthetic networks, including random Erd\H{o}s-R\'{e}nyi (ER) and Barab\'asi-Albert (BA) networks. In Sec. IV, we present the numerical simulations on some realistic networks. Finally, we give conclusions in Sec. V.

\section{SIR model with game-based self-quarantine}

We start by considering the SIR model on a contact network comprising of $N$ nodes and $E$ edges. In the standard SIR model, nodes in the network can be in three different states: susceptible, infected, or removed (representing death or immunization). Each susceptible node is infected with probability $\beta$ at each time step if it is connected to an infected node. Meanwhile, each infected node transitions to the removed state with probability $\mu$. In our model, we further incorporate game-based mechanism, assuming that the susceptibles would decide whether to self-isolate or not by weighing the gain and loss of quarantine. 

Specifically, we employ game theory and introduce two cost functions for each node $i$: $C_i^{e}$ and $C_i^q$, which denote the perceived cost of exposure and the perceived cost of quarantine, respectively. It is worth noticing that in reality, individuals may not evaluate these costs precisely. We assume that the perceived cost of exposure (or the perceived risk of infection) increases with the prevalence of an epidemic. In particular, both the local infection status (the fraction of immediate infected neighbors a node has, denoted by $\rho_i^{nn}$) and the global prevalence level (the fraction of infected nodes in the entire system, denoted by $\rho_I$) may contribute in $C_i^{e}$:

\begin{eqnarray}\label{eq:c1}
C_i^{e}=a_1 \rho_i^{nn} + a_2 \rho_I,
\end{eqnarray}
where $a_1, a_2 \in [0,\infty)$ are two independent control parameters, representing the sensitivity to local and global infection information, respectively. Larger $a_1$ (or $a_2$) indicates individuals are more responsive to the increase of infected nodes. The second term in Eq. (\ref{eq:c1}) acts like a mean-field effect, which exerts on all nodes. Note that $\rho_I$ (and $\rho_i^{nn}$) is time-dependent, thus $C_i^{e}$ is also time-dependent. On the other hand, we assume that the cost of quarantine for node $i$ is related to the degree $k_i$ of that node:
\begin{eqnarray}\label{eq:c2}
C_i^{q}=\eta(1-e^{-bk_i}),
\end{eqnarray}
where $\eta,b \in [0,\infty)$ are two independent parameters. In practice, $\eta$ should be adjusted to ensure that $C_i^{q}$ is comparable with $C_i^{e}$ (otherwise the model is meaningless). Without loss of generality, we set $\eta=1$ and $b=0.02$ in this paper. The above definition implies that the cost of quarantine increases with node degree, which is a very intuitive assumption, considering that severing each connection may incur a certain cost in reality. 

At each time step, every susceptible node decides to self-isolate and enters the quarantined state (denoted as $Q$) if $C_i^{e}>C_i^{q}$. Meanwhile, the quarantined nodes return to the susceptible state if $C_i^{e}<C_i^{q}$. In summary, our model can be represented as follows:
\begin{equation}\label{eq:seir}
\begin{array}{lc}
S_i+I_j \xrightarrow{\beta} I_i+I_j \\
I_i \xrightarrow{\mu} R_i \\
S_i \xrightarrow{C_i^{e}>C_i^{q}} Q_i \\
Q_i \xrightarrow{C_i^{e}<C_i^{q}} S_i.\\
\end{array}
\end{equation}

\begin{figure}
\epsfig{figure=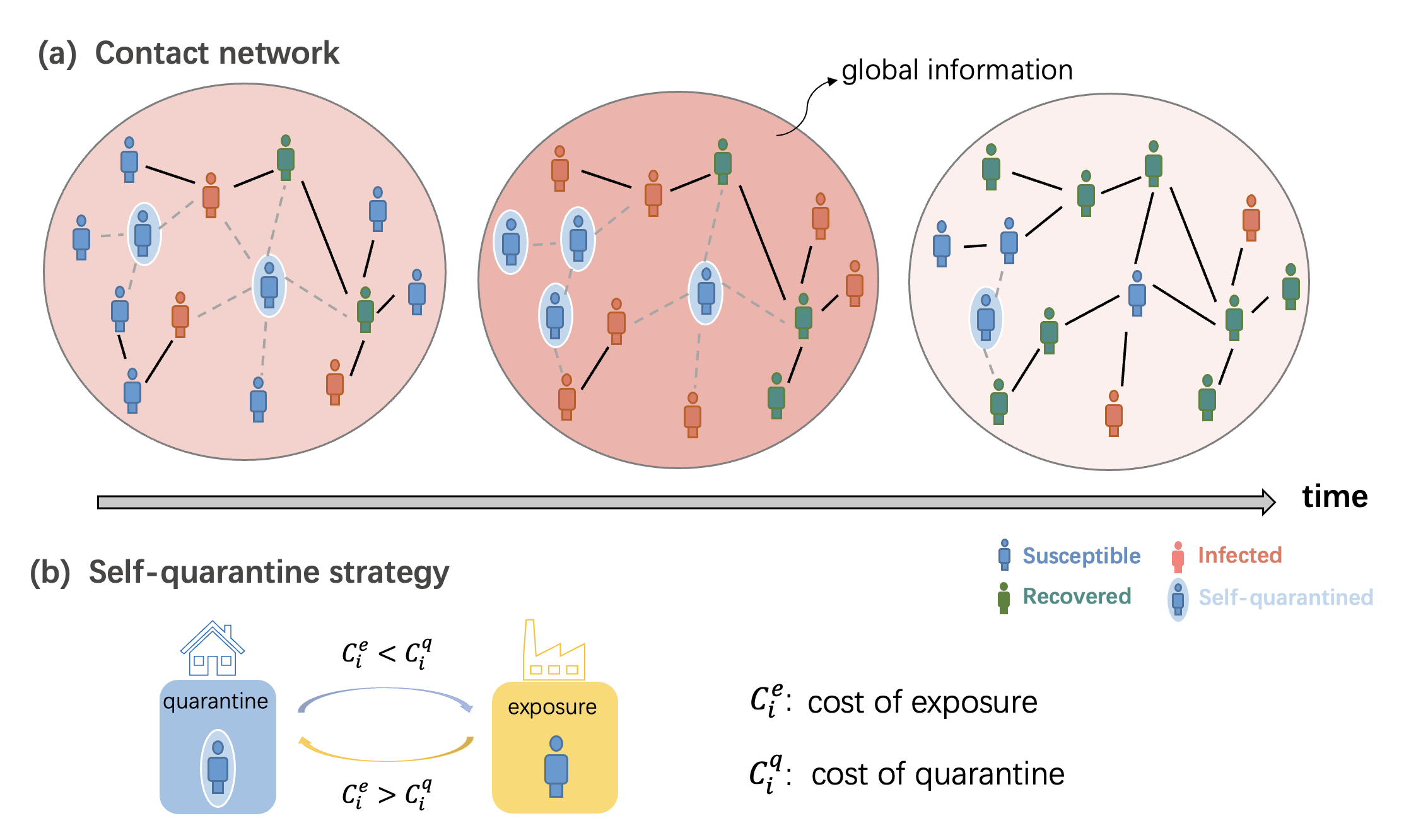,width=1.0\linewidth} \caption{(color online).  {\bf Schematic illustration of the epidemic model.} (a) Evolution of the contact network. The global information (related to the overall infected density $\rho_I$) acts like a mean field, which may exert influence over each susceptible node. The local infection status varies from one node to another. Note that the self-quarantined nodes cannot be infected by other nodes, and thus are equivalent to being removed from the network temporally.  (b) Self-quarantine strategy for nodes. If the cost of quarantine exceeds the cost of exposure, a node will decide not to self-isolate (or release from the quarantined state); Otherwise, it will enter the quarantined state.} \label{Fig:1}
\end{figure}

Note that the quarantined nodes are equivalent to being removed from the network temporarily \cite{Granger:2023,Bestehorn:2022}, which may rejoin the network (along with their links) through the game strategy. Our model is a generalization of the standard SIR model, and we refer to it as the SIR-Q model. The schematic illustration of the model is shown in Fig.\ref{Fig:1}.

\section{SIR-Q model in synthetic networks}
We first focus on a random ER network with size $N=5000$ and number of edges $E=25,000$. We put these edges randomly between each pair of nodes, which results in an exact average degree $\langle k\rangle=10$, and a Poisson distribution in node degrees (the detailed algorithm for generating an ER network is given in the supplementary material) \cite{Albert:2002}. The epidemic parameters are chosen as: $\beta=0.012$ and $\mu=0.01$. Initially, $0.5\%$ of nodes are selected at random as seeds to be infected, while all the remaining nodes are in the susceptible state. To quantify the spreading effect, we define $\rho_X$ ($X = S, I, R, Q$) as the fraction of the $X$ component among all nodes.

\begin{figure}
\epsfig{figure=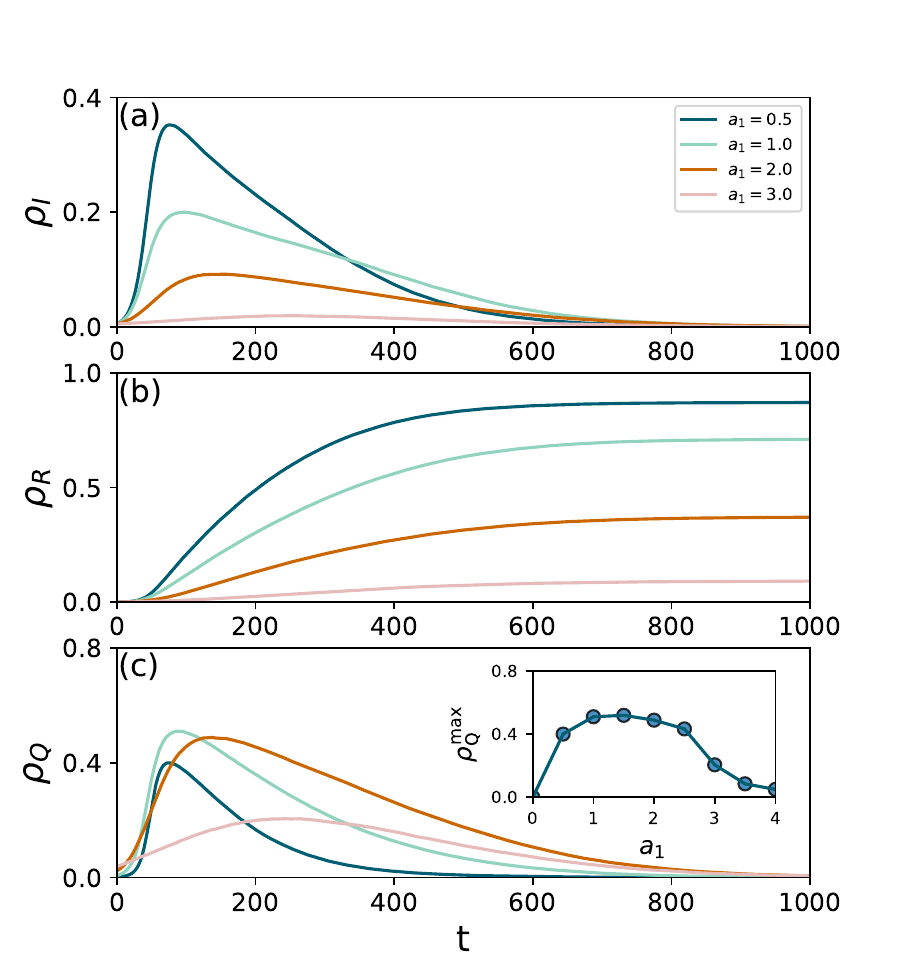,width=1.0\linewidth} \caption{(color online). {\bf Epidemic spreading under the influence of local information only.} Fraction of (a) infected nodes $\rho_I$, (b) removed nodes $\rho_R$, and (c) quarantined nodes $\rho_Q$ as a function of time $t$ for varying $a_1$ with fixed $a_2=0$. The inset in (c) shows how the maximum value of $\rho_Q$ changes with $a_1$. Curves are averaged over $100$ realizations.} \label{Fig:2}
\end{figure}

To proceed, we consider a simple case where $a_2=0$, meaning that individuals can only be affected by local information. Figure \ref{Fig:2} shows the fraction of nodes in different states ($\rho_I, \rho_R$ and $\rho_Q$) as a function of time $t$ for various $a_1$. We see that as $a_1$ increases, the peak of $\rho_I$ declines [as shown in Fig.  \ref{Fig:2} (a)], and the asymptotic value of $\rho_R$ decreases [as shown in Fig.  \ref{Fig:2} (b)], meaning that larger $a_1$ could effectively suppress the epidemic. This result is straightforward, since raising $a_1$ enhances the  exposure cost ($C_i^e$) of nodes connecting to infected individuals, thereby motivating them to self-isolate. Consequently, the spreading process is impeded.

On the other hand, the number of quarantined nodes displays a non-monotonic change with $a_1$. As illustrated in Fig. 2 (c), we observe that the peak of $\rho_Q$ rises as $a_1$ increases at first, primarily due to the growth in $C_i^e$ for nodes that are connected to infected individuals. However, further increasing $a_1$ leads to a decrease in $\rho_Q$. This behavior can be qualitatively understood as follows: For large values of $a_1$, the susceptible nodes become highly sensitive to their infected neighbors. As a result, the initial seeds could be effectively isolated by the quarantined nodes, making it difficult for the disease to spread. In this case,  only a very limited number of nodes (those connecting to the infected ones) experience an increase in their exposure cost $C_i^e$. For the majority of susceptible nodes, their exposure cost remains low (lower than the quarantine cost). Consequently, only a few nodes opt for self-isolation in the system. 

\begin{figure}
\epsfig{figure=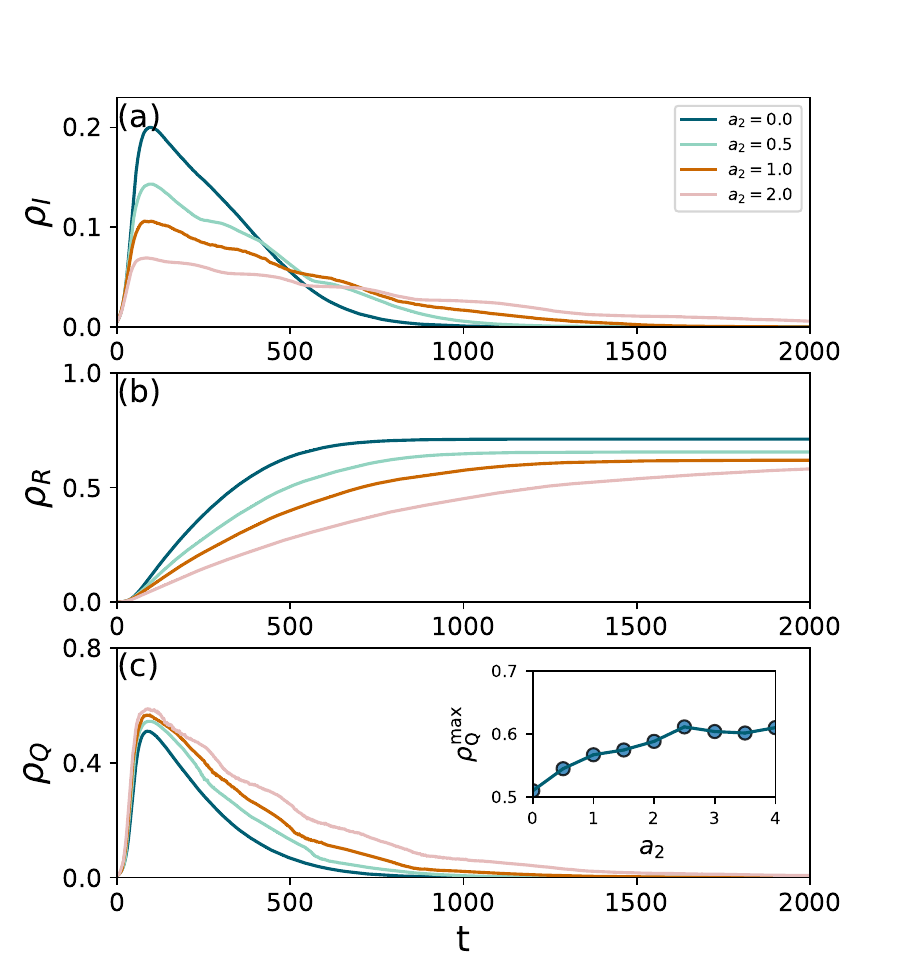,width=1.0\linewidth} \caption{(color online). {\bf Epidemic spreading under the influence of global information.} Fraction of (a) infected nodes $\rho_I$, (b) removed nodes $\rho_R$, and (c) quarantined nodes $\rho_Q$ as a function of time $t$ for varying $a_2$ with fixed $a_1=1$. The inset in (c) shows how the maximum value of $\rho_Q$ changes with $a_2$. Curves are averaged over $100$ realizations.} \label{Fig:3}
\end{figure}

Upon initial inspection, from Eq. (\ref{eq:c1}), it appears that global infection information ($\rho_I$) plays a similar role to local information ($\rho_i^{nn}$). Therefore, one might expect that the influence of parameter $a_2$ to be the same as $a_1$. Indeed, we observe that as $a_2$ increases, the epidemic spreading is suppressed, as shown in Fig. \ref{Fig:3} (a) and (b). However, we also note two distinct differences: 1. Epidemic curves exhibit shaking during the declining phase of the epidemic, as shown in Fig. \ref{Fig:3} (a). 2. The peak of the quarantined density increases monotonically with $a_2$ (and finally saturates), indicating that a significant number of nodes need to self-isolate during the spreading process to contain an epidemic. This second result is not difficult to comprehend if we notice that all susceptible nodes (even those without any infected neighbors) would be synchronously affected by the infected individuals at each time step.

\begin{figure}
\epsfig{figure=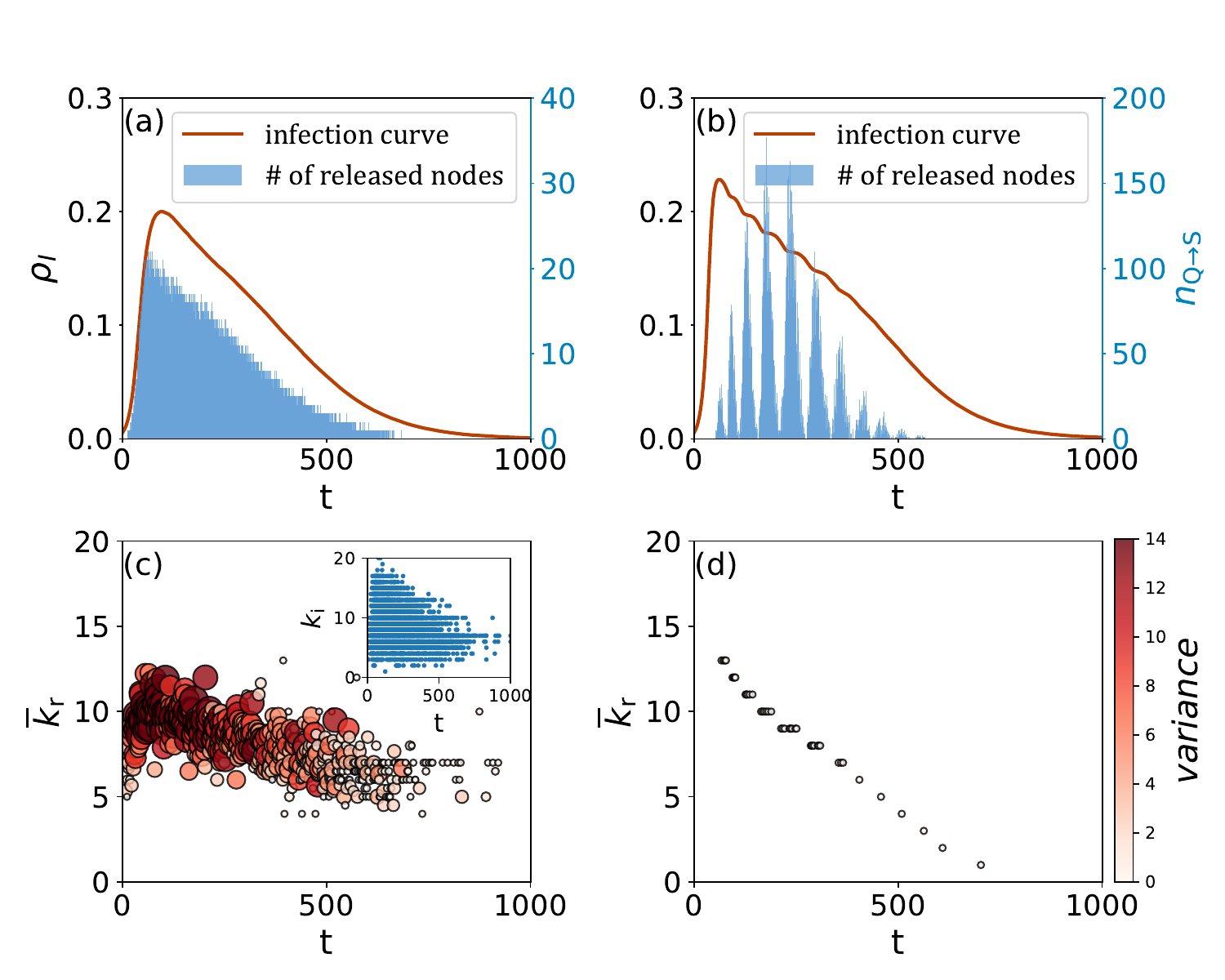,width=1.0\linewidth} \caption{(color online). {\bf The release pattern of nodes transitioning from quarantined state to susceptible state under the effect of local or global information.} Evolution of the fraction of infected nodes (the red curves) for (a) $a_1=1$, $a_2=0$, and (b) $a_1=0$, $a_2=1$. The number of nodes released from the quarantined state at each time step is shown as well (marked as blue bars). These results are averaged over $100$ realizations. Average degree of the released nodes $\bar k_r$ as a function of time $t$ for (c) $a_1=1$, $a_2=0$, and (d) $a_1=0$, $a_2=1$. The color (and the size of circles) represents the variance of degrees. The inset in (c) shows the detailed degree of each node released at every time $t$. These results are obtained from one single realization.} \label{Fig:4}
\end{figure}

To understand the shaking phenomenon, we calculate the number of nodes released from the quarantined state at each time step, denoted as $n_{Q\to S}$. Figure \ref{Fig:4} (a) and (b) depict the variation of $n_{Q\to S}$ with respect to time $t$ for $a_2=0$ ($a_1>0$) and $a_2>0$ ($a_1=0$), respectively. The corresponding evolution curves of the infected density $\rho_I$ are also presented (indicated by the red curves). It is evident that the processes of releasing nodes from quarantined state differ significantly between the two cases. In the scenario of $a_2=0$ (i.e., only local information is considered), $n_{Q\to S}$ decreases smoothly over time during the decay phase of the epidemic. While the inclusion of global information (corresponding to $a_2>0$) may result in ``bursty" behaviors. Notably, we observe that the quarantined nodes are released discretely (and concentratedly) in groups over time [the blue bars in Fig. \ref{Fig:4} (b)]. The concentrated emergence of susceptibles (from the pool of quarantined nodes) may lead to an extensive infection of nodes within a very short period of time, further contributing to the shaking behavior in the infection curves. 

To gain deeper insights, we further investigate the degrees of the released nodes at each time step. Define $\bar k_r(t)=\frac{1}{n_r(t)}\sum_{i \in \Theta_r(t)} k_i$ as the average degree of the nodes released at time $t$, where $k_i$ represents the degree of node $i$, $\Theta_r(t)$ is the set of the released nodes at time $t$, and $n_r(t)$ is the size of the set. Figure \ref{Fig:4} (c) and (d) present $\bar k_r$ as a function of time for the two different cases, corresponding to Fig. \ref{Fig:4} (a) and (b), respectively. It is noteworthy that $\bar k_r(t)$ displays a negative correlation with $t$ in both two cases, indicating that nodes with lower degrees tend to be released later in time. This can be attributed to the condition for releasing, i.e., $C_i^{e}<C_i^{q}$, which suggests that $k_i > -\frac{1}{b}ln(1-\frac{a_1}{\eta}\rho_i^{nn}-\frac{a_2}{\eta}\rho_I)$. Therefore, nodes with high degrees are more likely to fulfill this condition and would be released early. Nevertheless, due to the fluctuation of $\rho_i^{nn}$ across nodes (if $a_1 \ne 0$), nodes with various degrees may satisfy the above condition simultaneously, resulting in a large degree variance among the released nodes given time $t$ [see Fig. \ref{Fig:4} (c)]. In contrast, under the influence of global information (and $a_1=0$), the release of nodes occurs in a more regular manner --- nodes with the same degree would be released synchronously. This is indicated by a degree variance of $0$ among the nodes released at time $t$, as shown in Fig. \ref{Fig:4} (d). Moreover, since $\rho_I$ decreases continuously in the declining phase of the epidemic, nodes would be released successively in descending order of their degrees.

\begin{figure}
\epsfig{figure=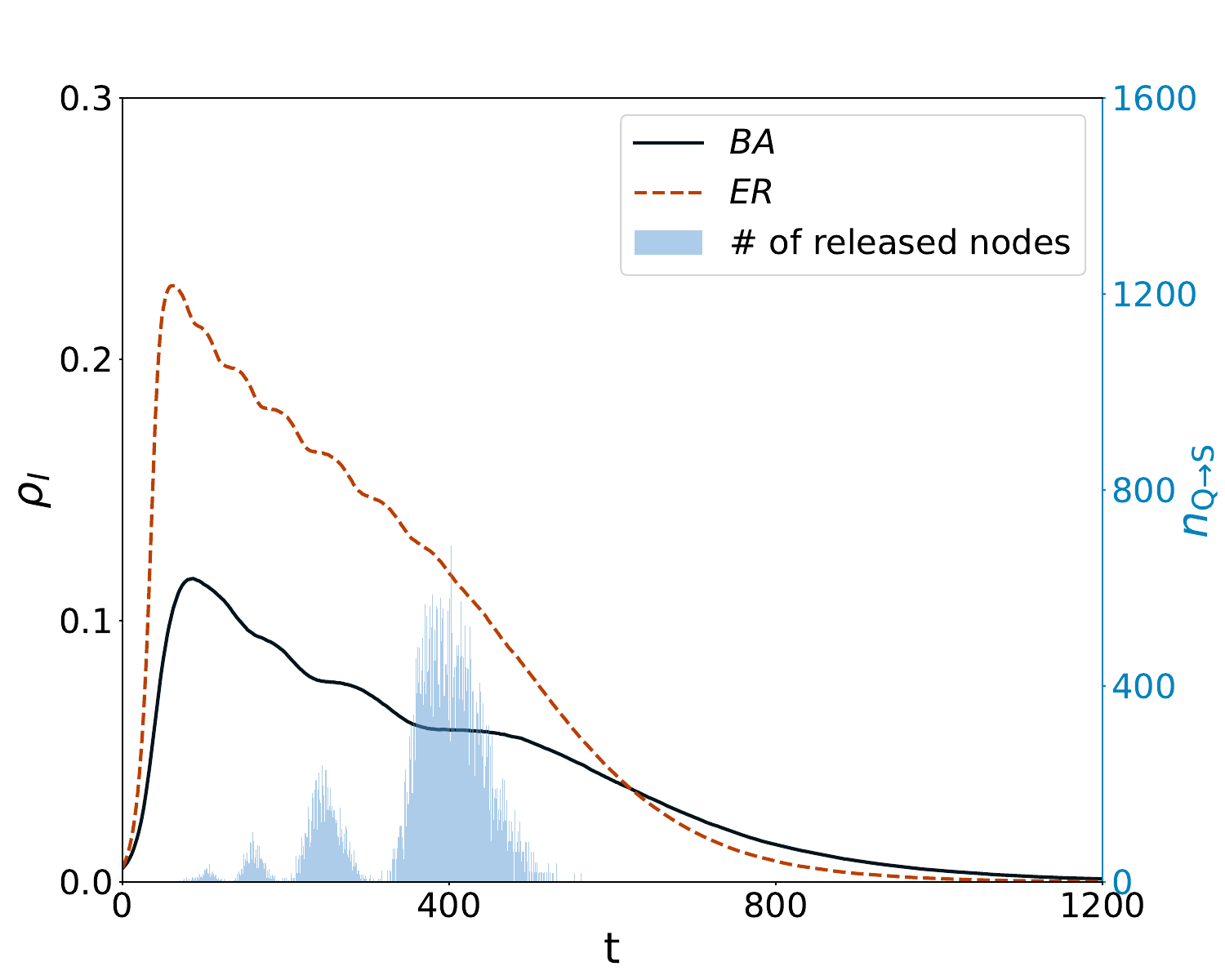,width=1.0\linewidth} \caption{(color online). {\bf Comparison of the infection curves in BA and ER Networks.} Fraction of infected nodes as a function of time $t$ in BA (the solid line) and ER networks (the dashed line). The blue bars correspond to the number of nodes released from the quarantined state at each time step in the BA network. In simulations, the size and the average degree of the two networks are kept the same, i.e., $N=5000$, and $\langle k\rangle=10$. The parameters $a_1=0$ and $a_2=1$. Results are averaged over $100$ realizations.} \label{Fig:5}
\end{figure}

The aforementioned shaking phenomenon is common and can also be observed in heterogeneous networks. To confirm this, we perform simulations on the BA model (the detailed algorithm for generating a BA network is given in the supplementary material) \cite{Albert:2002,Barabasi:1999}. In this network, the degree distribution follows a power-law distribution of the form $P(k) \sim k^{-3}$, indicating that the number of nodes with degree $k$ is a decreasing function of $k$. Figure \ref{Fig:5} shows the time evolution of $\rho_I$ and $n_{Q\to S}$ in the BA network. Again, we find that susceptible nodes are released discretely in groups. Different from the case of random ER networks, the group size increases with time in the BA network, owing to the delayed release of the large number of low-degree nodes. Correspondingly, the shaking behavior becomes increasingly notable as time progresses. To acquire an intuitive perception, we estimate roughly the number of nodes that are released at time $t$.  The lower boundary of degree that a node has such that it could be possibly released at time $t$ is $k_c(t) =\lceil  -\frac{1}{b}ln(1-\frac{1}{\eta}\rho_I(t)) \rceil$, where we have assumed $a_1=0$ and $a_2=1$. In the BA network, the number of nodes that satisfy this condition is $l(t)\sim \int_{k_c}^\infty k^{-3} dk = k_c(t)^{-2}$. During the declining phase of an epidemic (i.e., $\rho_I(t)$ decreases with $t$), $k_c(t)$ is a decreasing function of $t$, thereby $l(t)$ increases with time $t$.

Additionally, by comparing the infected density in the BA network with that in the ER network, we observe that the epidemic is significantly suppressed. This finding seems contradictory to the conventional understanding that heterogeneous networks can facilitate the spread of an epidemic \cite{Moreno:2002,Satorras:2001,Ruan:2020}. To understand this, it is important to notice that the previous studies were based on the assumption that most of nodes are susceptible, allowing hubs to effectively transmit the disease to a large portion of the network. However, in our model, the low-degree nodes are prone to self-isolate. Hence, in the BA network, which contains a large number of low-degree nodes, the epidemic is difficult to spread, even though the hubs may be readily contracted by the disease.

\section{SIR-Q model in real networks}

To further support the results obtained in the previous section, we analyze the SIR-Q model on some real-world networks. The first example is a directed friendship network obtained from a survey on adolescent health (AH) \cite{website1}. Each student participating in the survey was asked to list their top 5 female and top 5 male friends. In the network, each node represents a student, and a directed link from node $i$ to node $j$ indicates that student $i$ listed student $j$ as one of their best friends. We disregard the weight of these relationships in our analysis. The second example is a scientific collaboration network that encompasses papers on General Relativity and Quantum Cosmology (GrQc) submitted to the e-print arXiv between January $1993$ and April $2003$ \cite{website2}. In this network, each node represents an author, and if two authors co-authored a paper, an undirected edge is added between them. The third network is bipartite, involving sex buyers and their escorts (SE) \cite{Rossi:2015,Rocha:2010}. Specifically, in this graph nodes represent buyers and escorts, and each edge indicates sexual relationship between a buyer and an escort. The main properties of the three networks are summarized in Table \ref{tbl:table1}. %The third dataset pertains to dynamic contact networks, which were collected during the INFECTIOUS exhibition held at the Science Gallery (SG) in Dublin, Ireland, from April $17$th to July $17$th, $2009$ \cite{Isella:2011}. If two individuals were in contact for a minimum interval of $20$ seconds, they are linked as nodes in the network. By integrating the contact instances over the entire time window, we obtain an aggregated network, referred to as SG. 

\begin{table}[htb]
\renewcommand\arraystretch{1.2}
\centering
\caption{Statistics of the real-world networks. The parameters under consideration are: $N$, number of nodes; $E$, number of edges; $k_{max}$, maximum  degree; $\bar c$, average clustering coefficient \cite{Albert:2002}; $G$, Gini coefficient, which measures the skewness of the degree distribution \cite{Kunegis:2012}.}
\begin{tabular}{@{\hspace{0.3cm}}l@{\hspace{0.4cm}}@{\hspace{0.4cm}}c@{\hspace{0.4cm}}c@{\hspace{0.4cm}}c@{\hspace{0.4cm}}c@{\hspace{0.4cm}}c}
\toprule[0.5pt]
\toprule[0.5pt]
Data set & $N$ & $E$ & $k_{max}$ & $ \bar c $ & $G$   \\
  \midrule[0.5pt]
  AH & $2,539$ & $12,969$  & $36$ & $0.14$ &  $0.29$  \\
  GrQc & $5,242$ & $14,496$  & $81$ & $0.53$ &  $0.55$\\
%  SG & $10,972$ & $44,517$  & $64$ & 0.45 & $0.42$   \\
   SE & $10,106$ & $39,024$  & $313$ & 0.007 & $0.59$  \\
 
  \bottomrule[0.5pt]
  \bottomrule[0.5pt]
  \end{tabular}
  \label{tbl:table1}
\end{table}

\begin{figure}
\epsfig{figure=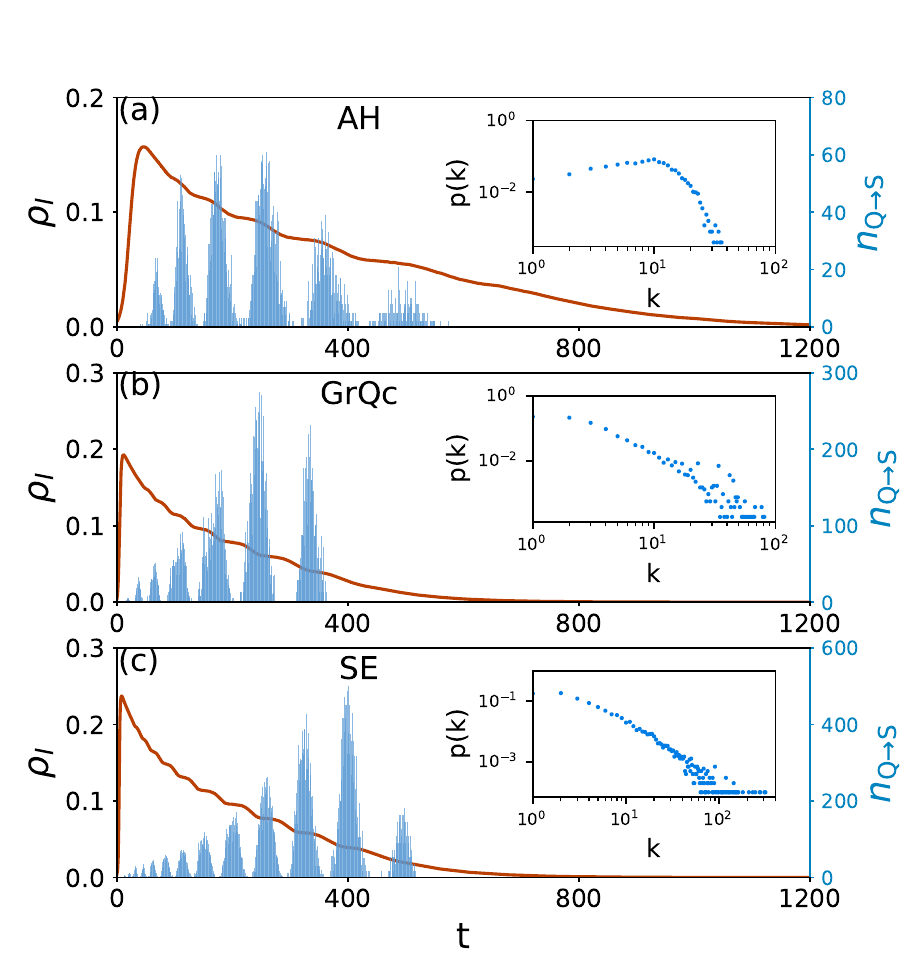,width=1.0\linewidth} \caption{(color online). {\bf Epidemic spreading in real networks.} Fraction of infected nodes as a function of time $t$ in real-world networks. The infection probability is chosen differently in simulations: (a) $\beta=0.04$, (b) $\beta=0.2$, and (c) $\beta=0.1$. The other parameters are given as: $a_1=0$, $a_2=1$, $\mu=0.01$. Curves are averaged over $100$ realizations.} \label{Fig:6}
\end{figure}

The degree distributions of the three networks mentioned above are presented in the insets of Fig. \ref{Fig:6}. We apply our epidemic model to each network and examine how the fraction of infected nodes $\rho_I$ changes with time $t$. In simulations, we set $a_1=0$ and $a_2=1$, i.e., only global information is available. As depicted in Fig. \ref{Fig:6} (a)-(c), it is evident that in each network, the evolution curve exhibit shaking during the declining phase of the epidemic. Furthermore, for each network, we present the number of nodes released from the quarantined state at each time step. As expected, we find that they are released by groups over time, which is consistent with the previous analysis. In particular, for the GrQc and SE networks, their degree distributions are more like a power-law, and correspondingly, the release patterns are more similar to that of the BA network, as well as the infection curves. Finally, It should be emphasized that these networks differ fundamentally in many aspects, which highlights the universality of our findings. 

\section{Conclusion} 

Our aim in this paper is to explore the distinct effects of local and global information on the spread of epidemics within the game-epidemic framework. Specifically, we proposed an extended SIR model on networks that incorporates individual self-quarantine behavior based on game strategies, assuming that individuals would make decisions whether to self-isolate or not by considering the perceived costs of infection and quarantine, guided by local and global information. We have shown that (i) enhancing the sensitivity of nodes to local information can effectively curb the epidemic by influencing a small number of nodes to self-isolate, while global information requires a more substantial number of nodes to engage in self-isolation; (ii) global infection information could trigger synchronous release of nodes with the same degree from the quarantine state, resulting in shaking in the epidemic evolution curves; (iii) in our model, the heterogeneous distribution of node degrees hinders the spread of a disease, as opposed to facilitating it, which is intrinsically different from traditional epidemic models. This model may provide valuable insights into understanding the interplay between epidemics and the prevalence information, as well as contribute to the epidemic control in reality (for example, more attention should be paid on low-degree nodes). 

Finally, it is essential to mention that the network models employed in our study are straightforward, neglecting some crucial features like the multi-layered nature of real systems and higher-order interactions among nodes. Further studies could integrate these features into the framework we proposed here. Furthermore, this framework may also find application in other domains, such as social contagions \cite{Granovetter1983,watts2002,ruan2015,Iniguez2018}. 

\section*{Supplementary Material}
See supplementary material for 1. Detailed algorithms for generating the ER and BA networks used in the main text; 2. The final epidemic size as a function of parameters $a_1$ and $a_2$; 3. The source code of the proposed model. 

\section*{Acknowledgement}

%We thank the SocioPatterns collaboration for providing the dynamical network data. 
This work was partially supported by the Key R\&D Programs of Zhejiang, China under Grant No. 2022C01018, by the Zhejiang Provincial Natural Science Foundation of China under Grants No. LR19F030001 and No. LY21F030017, and by the National Natural Science Foundation of China under Grant No. 61973273.

%\section*{Additional information}

%{\bf Competing financial interests}:
%The authors declare no competing financial interests.

\end{document}